\documentclass[twocolumn,secnumarabic,amssymb, nobibnotes, aps, prx, superscriptaddress]{revtex4-1}

\usepackage{soul}
\usepackage{graphicx}
\usepackage{graphics}
\usepackage{xcolor}
\usepackage[normalem]{ulem}
\usepackage{amsmath}
\usepackage{float}

\def\={\,=\,}

\begin{document}

%TC:ignore
\title{
%Noise-induced stabilization of  dynamical states in a non-Markovian system  
%Stabilizing Shapiro Steps in an RF Driven Josephson Junction with Noise
Noise-induced stabilization of dynamical states with broken time-reversal symmetry 
}

\author{T. F. Q. Larson} 
\affiliation{Department of Physics, Duke University, Durham, NC 27708, USA.}
\author{L. Zhao}   
\affiliation{Department of Physics, Duke University, Durham, NC 27708, USA.}
\author{E. G. Arnault}   
\affiliation{Department of Physics, Duke University, Durham, NC 27708, USA.}
\author{M. T. Wei}   
\affiliation{Department of Physics, Duke University, Durham, NC 27708, USA.}
\author{A. Seredinski} 
\affiliation{School of Sciences and Humanities, Wentworth Institute of Technology, Boston, MA 02115, USA.}
\affiliation{Department of Physics, Duke University, Durham, NC 27708, USA.}
\author{H. Li}   
\affiliation{Department of Physics and Astronomy, Appalachian State University, Boone, NC 28607, USA.}
\author{K. Watanabe}
\author{T. Taniguchi}
\affiliation{Advanced Materials Laboratory, National Institute for Materials Science, 1-1 Namiki, Tsukuba, 305-0044, Japan.}
\author{F. Amet} 
\affiliation{Department of Physics and Astronomy, Appalachian State University, Boone, NC 28607, USA.}
\author{G. Finkelstein}
\affiliation{Department of Physics, Duke University, Durham, NC 27708, USA.}

\date{\today}%

\begin{abstract} 
Under a high frequency drive, Josephson junctions demonstrate “Shapiro steps” of quantized voltage. These are dynamically stabilized states, in which the phase across the junction locks to the external drive. We explore the stochastic switching between two symmetric steps at $\frac{\hbar \omega}{2e}$ and $-\frac{\hbar \omega}{2e}$. %at zero external bias. 
Surprisingly, the switching rate exhibits a pronounced non-monotonicity as a function of temperature, violating the general expectation that transitions should become faster with temperature. We explain this behavior by realizing 
that the system retains memory of the dynamic state from which it is switching, thereby breaking the conventional simplifying assumptions about separations of time scales.
\end{abstract}
%TC:endignore
\maketitle

\newpage

%\section{Introduction}	
Nanoscale systems have recently emerged as a platform to study the stochastic effects and quantum thermodynamics~\cite{RMP09, pekola_towards_2015}. Such systems are often very tunable, providing high quality data across a wide range of parameters and system bandwidths, and allowing for the study of both classical and quantum fluctuations.  Several recent works have utilized quantum dots and single electron transistors tuned between charge states in order to achieve the desired bistability~\cite{koski_experimental_2014,wagner_quantum_2019} while other works have focused on superconducting quantum circuits~\cite{Masuyama2018, PhysRevLett.121.030604}. 
Here, we investigate switching in an RF-driven Josephson junction, where bistability is achieved between  
%phase space trajectories, which are equivalent to each other under time reversal, and corresponding to states which are \emph{dynamic} in nature. 
two dynamic states which are equivalent to each other under the time reversal.

Josephson junctions subject to an RF radiation demonstrate an inverse AC Josephson effect: the phase difference across a junction locks to the external frequency~\cite{JosephsonRMP1964}. As a result, the phase steadily ramps with time, changing by integer $n$ periods per RF cycle. The $I-V$ curves of the junction show ``Shapiro steps'' of quantized voltage, $V= \frac{n \hbar \omega}{2e}$~\cite{Shapiro1963}. The extremely precise voltage quantization of the steps is presently utilized in primary voltage standards~\cite{hamiltonJosephsonVoltageStandards2000a}. The exact mechanisms and stability of the phase locking were investigated in detail in the 1980's, with an emphasis on stability of closed trajectories in the phase space~\cite{KautzRev1996}.

In this work, we study stochastic switching between these phase locked dynamical states in a gate-tunable graphene Josephson junction.  Stochastic switching between Shapiro steps has recently been observed in this system, with switching times exceeding the time scale of the dynamics by many orders of magnitude~\cite{PhysRevResearch.2.023093}. A slight change of the RF amplitude changes the switching time over an exponential range, from milliseconds to hours. The tunability and degree of control enabled by the Josephson junctions make them particularly appealing to study the switching dynamics in a driven-dissipative system.

\begin{figure}[h!]
    \centering
    \includegraphics{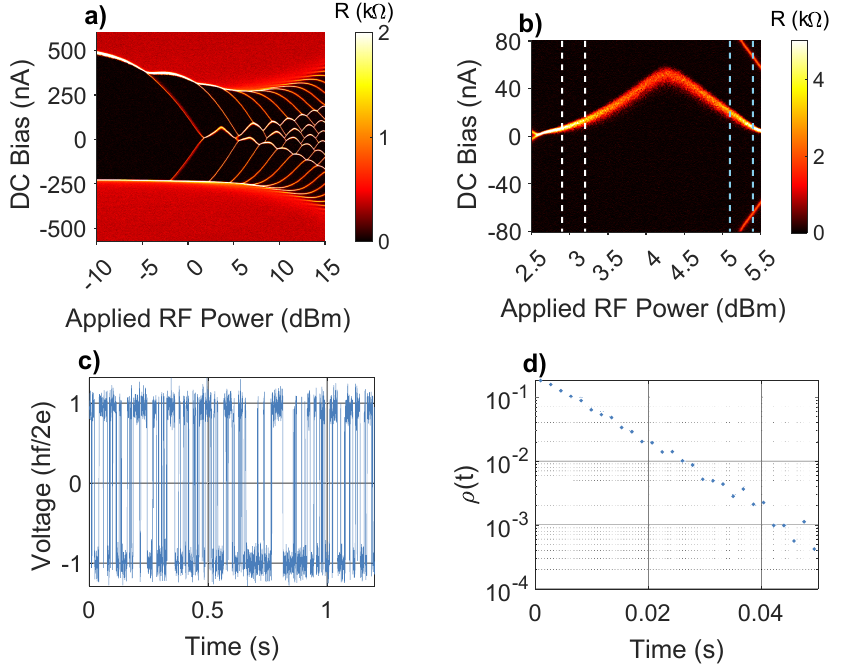}
\caption{a) A map of the differential resistance vs RF power and DC bias current. Dark regions correspond to the quantized steps, while the transitions between the steps appear as bright lines. The current is swept from negative to positive, and many transition boundaries are distorted by the long switching times. b) Zoom into the region in which the system has two stable quantized steps, $+1$ and $-1$, which are equally probable at zero DC bias. The bright feature corresponds to the transition between them, extracted by numerically differentiating an average of 200 current sweeps. In the following, we focus on a small range of the applied power corresponding to the region between the dashed white lines. c) A sample of the output voltage measured at 2.9 dBm over a small interval of time. The average time between switching events $\tau_0$ was extracted from these and similar traces, and care was taken to operate in the regime where the distribution of measured voltages remained bimodal. d) The probability distribution of switching times, taken from the full two minute trace corresponding to (c).} 
\end{figure}

We have previously showed an unusual pattern of Shapiro steps in graphene Josephson junctions~\cite{Larson2020}, rather different from the conventional Bessel function behavior~\cite{Tinkham}. 
Notably, at some intermediate RF power both $n=+1$ and $n=-1$ are stable solutions at zero current, while $n=0$ is not, resulting in ``zero-crossing steps'' in the $I-V$ curves ~\cite{KautzRev1996,PhysRevResearch.2.023093,Larson2020}. 

In the tilted washboard picture commonly used to describe Josephson junctions, zero current corresponds to zero average tilt of the the washboard. The $n=\pm 1$ states correspond to the ``phase particle'' ratcheting one period per RF cycle either forward or backward along the washboard, therefore representing a peculiar example of a spontaneously broken time reversal symmetry in a driven-dissipative system. Over the course of many drive cycles, rare fluctuations knock the phase particle so that it switches the direction of motion. 
Here, we find that the temperature dependence of this switching lifetime shows a striking non-monotonicity. Concomitantly, the step voltage gets reduced compared to the quantized step value. Accounting for these features necessitates non-trivial phase space dynamics occurring at time scales much slower than the RF oscilations. 

The Josephson junctions studied in this paper are made from graphene encapsulated in hexagonal Boron Nitride. The superconducting contacts are made from MoRe alloy ($T_{C} \approx 10$ K). Such junctions are gate tunable and mediate supercurrent via Andreev bound states~\cite{Tinkham}. However we believe the physics presented in this work does not depend on the microscopic properties of the sample and could be realized in other types of Josephson junctions with suitably similar parameters. The measurements are conducted in a dilution refrigerator with a microwave antenna placed close to the sample for AC biasing. The RF frequency is 5.0 GHz, and all figures except for Figure 4 which is measured on a different device at 3.2 GHz. % measured in the sample junction.  \GF{Are you stressing 5 vs 5.2? Or was this a leftover from current Fig. S5?}\TL{Stressing 5.2 and different sample}

In Figure 1a, we reproduce the map of differential resistance vs bias and RF power studied in Ref.~\cite{Larson2020}. The dark regions correspond to Shapiro steps, where $dV/dI$ vanishes. The map is measured by sweeping bias $I_{DC}$ from negative to positive. The hysteresis in switching between the plateaus results in the pronounced top-bottom asymmetry of the map, noticeable as arching of the transition boundaries. Sweeping the current bias in the opposite direction would flip this map about the horizontal axis.

Figure 1b shows a zoom-in map of the region in which the system switches between the $n=-1$ and $+1$ steps. To obtain both maps, multiple $I-V$ curves are measured by sweeping the bias $I_{DC}$; these curves are then averaged and numerically differentiated. Therefore, the bright band of high $dV/dI$ in the middle of the map corresponds to the average value of the switching current. Furthermore, the cross-section of the band represents the histogram of the switching currents for the positive sweep direction. As in Figure 1a, sweeping the current bias in the opposite direction would flip this map vertically. 

For the RF powers in the middle of Figure 1b ($\sim 4.2$ dBm), the $n=\pm 1$ states at $I_{DC}=0$ could persist for hours. The lifetimes are much shorter at the edges of the region, allowing us to record multiple transitions between the quantized steps while holding all control parameters constant. The primary region studied in this paper is just above the first bifurcation point, $P_{RF} \approx 2.6$ dBm, at which the $n=0$ state first becomes unstable and $n=\pm 1$ states are observed. The studied range is marked by the dashed white lines in Figure 1b. 

In Figure 1c, we plot a portion of a typical time trace measured at zero bias and $P_{RF}=2.9$ dB. The voltage stochastically switches between the two plateaus with an average time of $\tau_0\equiv \langle \tau \rangle \approx 10$ ms. The probability for a step to last a given time $\rho (\tau)$, as observed over the full time trace, is plotted in Figure 1d, demonstrating a clear exponential dependence with a slope of $\log (\rho) = 1/\langle \tau \rangle$ as expected for uncorrelated random processes. 

\begin{figure}[t]
    \includegraphics{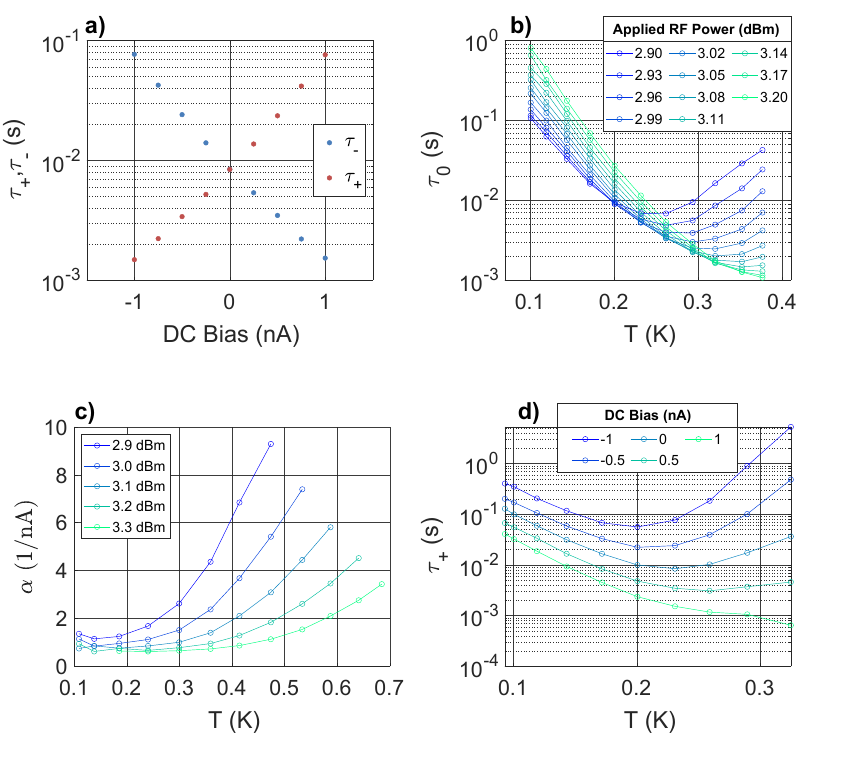}
    \caption{ a) Switching times from $n=-1$ to $n=+1$ and back, $\tau_+$ and $\tau_-$, vs current bias. The rates are exponentially sensitive to bias, and 
    $\alpha$, the slope of $\log (\tau_{\pm}/\tau_0)$ vs. $I$ is further plotted in panel (c).
    At zero bias, the two times are equal to each other and correspond to $\tau_0$ of Figure 1d. b) Dependence of $\tau_0$ on temperature at several values of $P_{RF}$ in the range marked in Figure 1b. While the expected steady decrease of lifetime with temperature is observed at higher power, $\tau_0(T)$ is non-monotonic for lower powers. It is important that the raw voltage traces show two well defined voltage states for all temperatures and powers shown.
    c) Slope $\alpha$ extracted from panel (a) and similar data, plotted vs $T$ for several applied powers. While at low temperatures the values of $\alpha$ are similar across applied powers and show relatively small temperature dependence, elevated temperatures show a rapid increase of $\alpha$, particularly at lower powers. d) Temperature dependence of $\tau_{-}$ at several values of bias $I$. Note that in the low temperature regime, where $\tau_0$ decreases with temperature, the $\tau_{-}(T)$ curves are parallel indicating a roughly constant $\alpha$. The curves start to diverge at higher temperatures, where $\tau_0$ increases with temperature and $\alpha$ grows with $T$.
    %\GF{pixelated bitmap, atrocious label of panel c, wrong labels (fixed)}
    } 
\end{figure}

Despite the fact that the two states of the system are dynamical in nature, it remains possible to consider this system as a double well problem in an effective ``quasipotential"~\cite{KautzRev1996, Chan2007,Ricci2017}. The double well potential can be tuned with DC bias which can be observed in Figure 2a. At zero bias, the quasipotential double well is very close to symmetric, with the lifetime of each state lasting an equal length of time.  By applying a small bias, on the order of 1 nA, it is possible to break the symmetry, shifting the imbalance in favor of one of the two states. Both times follow a clear exponential dependence $\log (\tau_{\pm}/\tau_0) = \pm \alpha I$, consistent with bias linearly increasing the activation gap. As a result, the product $\tau_{+} \tau_{-}$ remains nearly constant as a function of bias.

While all the features are so far consistent with a standard double well potential problem, the temperature dependence of $\tau$ is surprisingly different. This is shown in Figure 2b for temperatures from 100 mK to 400 mK and for a range of powers identified in Figure 1b. Considering first the power range $P_{RF}>3.1$ dBm (green curves), $\tau_0$ behaves as might be expected: it increases upon decreasing $T$ and moving away from the bifurcation point, qualitatively consistent with thermal activation. However, for lower powers, $P_{RF}<3.1$ dBm (blue curves), $\tau_0$ develops a peculiar non-monotonic behavior, where it \emph{increases} at higher temperatures. 

We have excluded possible trivial explanations for the observed effect. First, the measured voltages traces remain distinctly bimodal with two well-defined voltage states throughout the full temperature range. Second, for applied power just below the second bifurcation point (located at $P=5.1$ dBm), the $\tau_0(T)$ curves show a qualitatively similar behavior to Figure 2a (supplemental material). Namely the non-monotonicity moves to \emph{lower} temperatures as the system is tuned closer to the bifurcation point, which here corresponds to \emph{higher applied RF power.} We believe this excludes the possibility that non-monotonicity of $\tau_0(T)$ arises from a change of the system parameters with temperature. For example, a gradual reduction of $I_{C}$ would reduce the power corresponding to both bifurcation points. Therefore, for a fixed applied power, the lower bifurcation point would move away, while the upper bifurcation point would move closer, resulting in different behaviours in their vicinity.
We return back to the surprisingly non-monotonic $\tau_0(T)$ curves later in the text.
%For example, a gradual reduction of IC would cause the non-monotonicity to arise at lower powers close both bifurcation points. 

%\GF{This paragraph is bad (non-intuitive) and collectively this and the next paragraph break the flow.} \TL{

Noting this departure from predicted Arrhenius behavior, we next explore the combined effect of bias current and temperature on $\tau_{0}$. In Figure 2c, we plot the dependence of the slope $\alpha = \frac{d\ln \tau_+}{d I}$ on temperature at several values of $P_{RF}$. While at low temperatures $\alpha$ saturates at roughly the same level for different $P_{RF}$, at higher temperatures $\alpha$ rapidly {\emph increases}. Note that the picture of a particle in a bistable potential tilted by current would instead predict a $\alpha \propto 1/T$ dependence. 

For a given RF power, $\alpha$ begins to increase at temperatures just below the onset of non-monotonic behavior of $\tau_0$. The interplay between $\tau(T)$ and $\alpha(T)$ is best seen in Figure 2d, which plots $\tau_{+}$ vs. $T$ at $P_{RF}=2.9$ dBm for 5 values of bias. The middle curve corresponds to zero bias, at which $\tau_{+}=\tau_0$. At low $T$, where $\tau_{0}$ decreases with $T$, all curves are roughly parallel to each other (on a log scale), indicating a constant $\alpha$. However, at high temperatures they start to diverge, which indicates that $\alpha$ is no longer saturated; this occurs just before $\tau_0$ starts to grow with $T$. 

It has been shown through both analytical studies and numerical simulations~\cite{KautzRev1996}, that the transition rate between the two distinct steady states in the traditional RCSJ model is expected to follow an activation temperature dependence, with the activation energy $E_A$ determined by the the most probable path in the phase space connecting the two states.  Clearly, the non-monotonic dependence of $\tau_0(T)$ and the increase of $\alpha$ with temperature cannot be explained by such activation.

\begin{figure}[t]
    \includegraphics{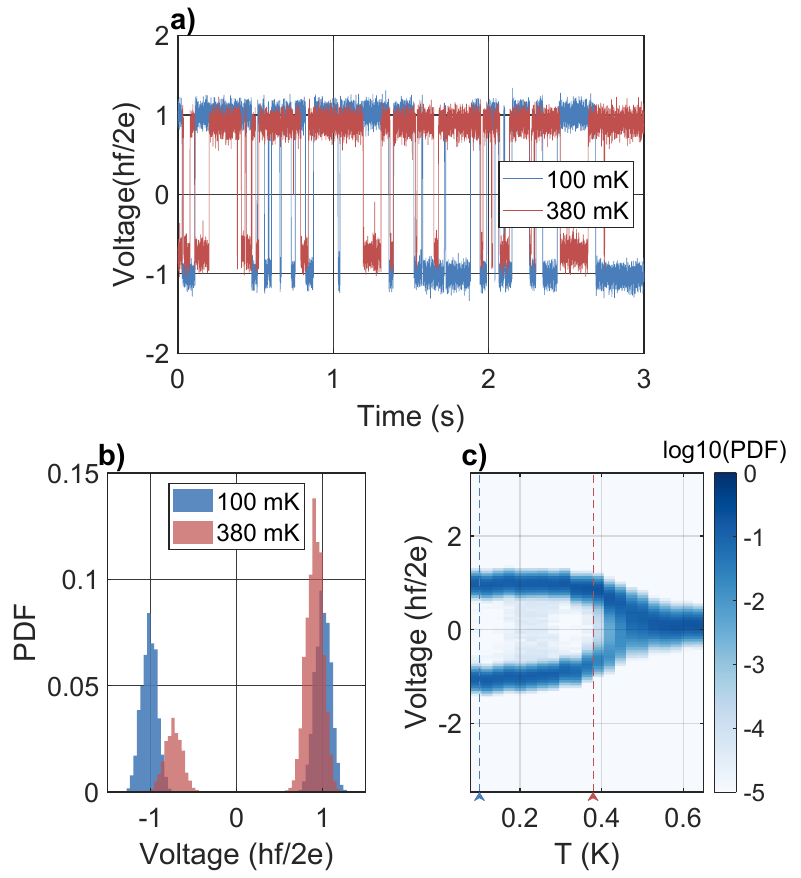}
\caption{ a) Time traces measured at $P_{RF}= 2.9$ dBm and temperatures of 110 mK (blue) and 380 mK (orange), which have similar switching times (see Figure 2b). b) The probability distribution of measured voltages corresponding to these traces. While the two steps are still well-defined, the 380 mK trace clearly shows the loss of quantization. c) Voltage distributions at multiple temperatures represented as a map. High probability region (blue) shows two peaks at $\pm \frac{hf}{2e}$ that eventually merge into one peak at $V=0$. 
}
\end{figure}

Inspection of the measured voltage traces in the non-monotonic $T$ regime reveals a notable anomaly: While the voltage still switches between distinct positive and negative steps, their values become significantly reduced from $\frac{\hbar \omega}{2e}$. 
One such trace is presented in Figure 3a (380 mK, blue dots). For comparison, we also present a trace at a lower temperature, selected such that the average switching times of the two traces are comparable (red dots, 100 mK). 
%One such trace, as well as a low temperature trace, are presented in Figure 3a for comparison. Here, the temperatures are selected such that the average switching times are comparable. 
Figure 3b shows the corresponding histograms of the measured voltages, which clearly demonstrate that at high temperature the average voltage corresponding the $n=\pm 1$ states is reduced.

We further demonstrate this effect in the map of Figure 3c, which is made of the voltage histograms similar to panel (b) taken at different temperatures. The resulting ``pitchfork" feature is non-trivial and very different from the expectations based on a conventional bistable system~\cite{singh_distribution_2016}. % as described below. %the thermal noise is 
There, in the temperature range where the switching rate is comparable to the measurement bandwidth, the distribution function forms a broad peak at zero voltage, which coexists with with the narrow peaks which stay at $n=\pm 1$~\cite{singh_distribution_2016}. 
As the temperature increases and switching becomes faster, the broad peak should grow while the peaks at $n=\pm 1$ should decrease in height and eventually disappear. We observe a very different picture: the two side peaks are shifting towards each other, before eventually merging to form the central peak.

%\sout{We argue below that the $n=\pm 1$ states still survive, but the system rapidly spikes from e.g. $-1$ to $+1$, before returning back to $-1$. If these spikes are too fast to be experimentally resolved, the average voltage of both states would appear to be reduced. }

The clue to this behavior has been obtained from a measurement of the second device under a 3.2 GHz RF drive (Figure 4). The different microwave frequency changes the depth of the effective potential and the switching rates, allowing us to observe fast events which were masked in Figure 3. 
Figure 4a shows a 600 second time trace (blue) which demonstrates very rapid switching at a rate of at least $1/\tau \geq 200$ Hz. Surprisingly, by numerically smoothing the signal to remove the high frequency component, we recover a bistable behavior with a very slow switching between two well-defined states, which are no longer quantized (red curve). Figure 4b plots a 60 ms segment of the raw trace right around one of the switching points of the averaged signal. It is clear that there are two regions with different averaged behavior: on the left, the system mostly stays around $-1$, while on the right if mostly stays close to $+1$. In both cases, there are many spikes reaching to the opposite state, which rapidly return back. We argue that the reduced voltage steps in Figure 3 are formed by the $n=\pm 1$ states with rapid spikes which are too fast to be resolved within our bandwidth. Indeed, the histogram of the time-filtered (red) curve in Figure 4c is similar to those found in Figure 3.

\begin{figure}[t]
    \includegraphics{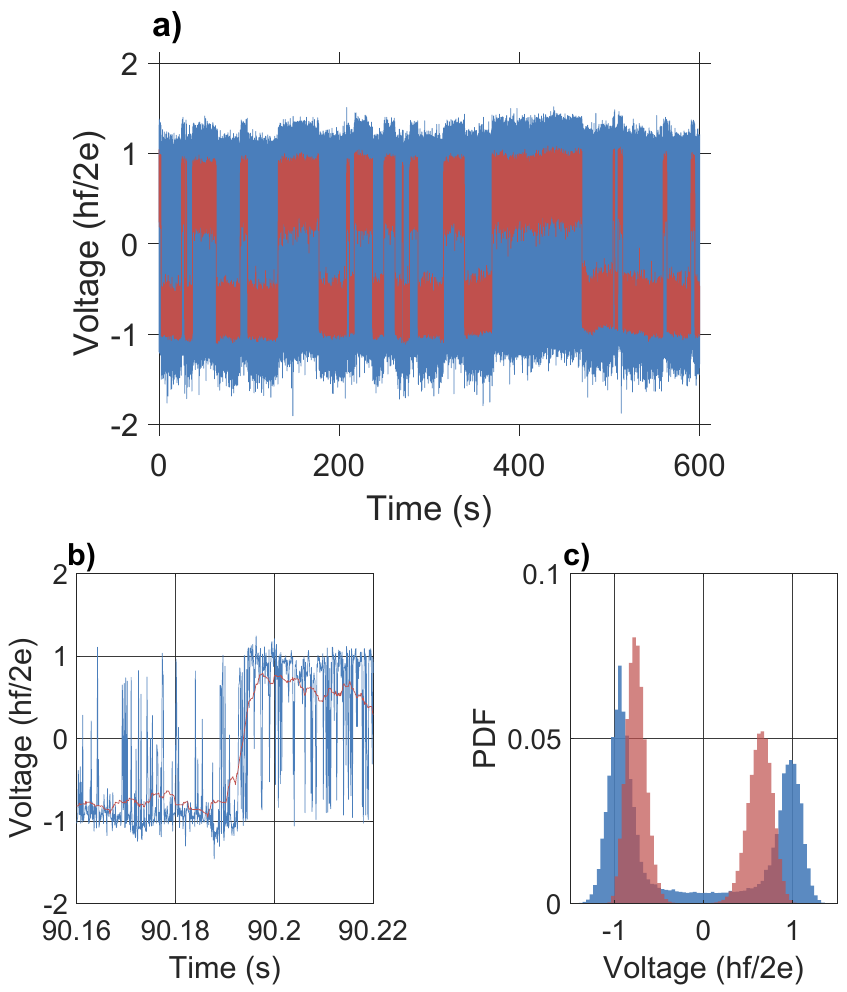}
\caption{Measurements of a second, similar device at $f=3.2$ GHz.
a,b) Blue line: the trace of the measured voltage as a function of time on a large scale (a) and zoomed in (b). The trace in shows many very short spikes -- likely aborted switching attempts -- after which the system returns back to the original step, as seen in panel (b). These spikes are quite different from the bistable behavior demonstrated in Figure 1d. The red line in panels (a) shows the same trace with higher frequencies numerically filtered out, which suppresses the spikes and restores the bistable behavior. The resulting trace has only a few successful switching events, one of which is centered in (b). c) The histograms of the traces in (a). The histogram of the original trace has two main peaks centred at the $\pm 1$ steps (blue). However there is also a significant weight in the region between the peaks. After the numerical filtering of the trace, the distribution becomes perfectly bimodal, but the peaks shift away from the quantized values (red). The latter histogram is similar to the ones found at elevated temperatures in Figure~3b.  
}
\end{figure} 

Our observations in Figure 4 indicate that while the smoothed data in panel (a) appears to be Markovian, the originally rapidly switching data is certainly not. Instead, the system has memory: the switching probability depends on how long the system spent in a given state. We believe this memory effect arises because of additional poles in our measurement circuit. Physically, the capacitance of the cryogenic RC filters and cabling is charged by the DC voltage corresponding to the Shapiro step. When the system attempts to switch between the dynamical states, e.g. from $-1$ to $+1$, the charged capacitors discharge through the junction, biasing it back to the $-1$ state. At low temperature, when the escape rate is low, this current will typically decay before the system switches back to the original state. But at higher temperatures, the system has a higher probability to return from $+1$ back to $-1$ while the capacitors have not yet discharged. The additional RC stages therefore stabilize both $\pm 1$ states, leading to the highly correlated switching (spikes) observed in Figure 4b. This effect may be viewed as a form of noise enhanced metastability~\cite{Spagnolo_rev}, although the specific model, and the enhancement in a dynamical bistability are both novel.

To quantitatively verify this picture, we have performed simple simulation which qualitatively reproduces the observed behaviors, including the correlated spikes of Figure 4 and the original non-monotonic temperature dependence of Figure 2 (supplementary Figure S6). One important question which remains open is the conceptual importance of the $n=0$ state in explaining the non-monotonic $\tau_0(T)$. Numerical simulations of the full differential equation typically show no direct transitions between the $+1$ and $-1$ states; instead the system transitions by passing through $n=0$. 
Experimentally, whenever the $0$ state is visible within our time resolution, the system tends to switch between  $-1$ and $+1$ via that state 
(supplementary Figure S5). Therefore, the triple well potential is both expected in our system and makes it easier to account for the nonmonotonicity of $\tau_0(T)$. However, the possibility remains that the memory effects could cause non-monotonic temperature behavior in a double well ($n=\pm 1$) even without the metastable $n=0$ state. 

In light of these observations, we could also speculate about the nature of the temperature dependence of the slope $\alpha$, shown in Figure~2c. At low temperatures, while $\tau_0$ is decreasing, the slope is determined by the transition rate between $-1$ and $0$, which appears to be relatively insensitive to bias. At higher temperatures, as the system becomes more sensitive to the transitions from $0$ to $\pm 1$, the slope becomes high, indicating that the bias has larger effect on the relatively shallow barriers separating $0$ from the other two state. This picture is qualitatively confirmed by the toy model presented in the Supplementary (Section S7).

To summarize, we have studied switching in an AC driven Josephson junction, which demonstrates surprisingly complex dynamical behavior. We focus on the regime with ``zero-crossing steps'', in which the zero voltage state is not stable at zero bias, and the system spontaneously develops a quantized voltage of $\pm \hbar \omega /2e$. The switching time between these states vary from $10^{-3}$ to $10^3$ seconds. We found unexpected non-monotonic temperature dependence of the switching time, which reaches a minimum value at some intermediate temperature. We attribute this behavior to the non-Markovian effects induced by the sample environment, which retains memory of the previous state of the system for a long time. These effects are combined with the complex structure of the phase space, in particular with the presence the metastable zero-voltage state. While we understand the overall features of the data, the nature of the most probable escape path, both in the Markovian and non-Markovian limits, is still an open question for this system.

The type of samples as studied here provide a flexible and highly tunable platform to probe the unexplored aspects of a driven-dissipative dynamics.  While the electron temperature in our system is likely too high for the phase particle to behave quantum mechanically~\cite{DeCecco2016}, it may be possible to engineer a device which undergoes such a dynamical bistability in the macroscopic quantum tunneling regime~\cite{Clarke1988}, opening doors to many interesting experiments on driven-dissipative quantum dynamics.
\newline
\newline
%TC:ignore
\acknowledgements
%\begin{acknowledgements}
We thank G. Refael, S. Teitsworth and especially M. Dykman for helpful discussions. Measurements of graphene samples by T.F.Q.L. and E.G.A, and data analysis by T.F.Q.L., L.Z. and G.F., were supported by Division of Materials Sciences and Engineering, Office of Basic Energy Sciences, U.S. Department of Energy, under Award No. DE-SC0002765.  Lithographic fabrication and characterization of the samples performed by M.T.W. and A.S. were supported by the NSF Award DMR-2004870. H. Li, and F.A. acknowledge the ARO under Award W911NF-16-1-0132.  K.W. and T.T. acknowledge support from JSPS KAKENHI Grant Number JP15K21722 and the Elemental Strategy Initiative conducted by the MEXT, Japan. T.T. acknowledges support from JSPS Grant-in-Aid for Scientific Research A (No. 26248061) and JSPS Innovative Areas “Nano Informatics” (No. 25106006). This work was performed in part at the Duke University Shared Materials Instrumentation Facility (SMIF), a member of the North Carolina Research Triangle Nanotechnology Network (RTNN), which is supported by the National Science Foundation (Grant ECCS-1542015) as part of the National Nanotechnology Coordinated Infrastructure (NNCI). 
%\end{acknowledgements}

\bibliography{main.bib}	

\begin{thebibliography}{20}%
\makeatletter
\providecommand \@ifxundefined [1]{%
 \@ifx{#1\undefined}
}%
\providecommand \@ifnum [1]{%
 \ifnum #1\expandafter \@firstoftwo
 \else \expandafter \@secondoftwo
 \fi
}%
\providecommand \@ifx [1]{%
 \ifx #1\expandafter \@firstoftwo
 \else \expandafter \@secondoftwo
 \fi
}%
\providecommand \natexlab [1]{#1}%
\providecommand \enquote  [1]{``#1''}%
\providecommand \bibnamefont  [1]{#1}%
\providecommand \bibfnamefont [1]{#1}%
\providecommand \citenamefont [1]{#1}%
\providecommand \href@noop [0]{\@secondoftwo}%
\providecommand \href [0]{\begingroup \@sanitize@url \@href}%
\providecommand \@href[1]{\@@startlink{#1}\@@href}%
\providecommand \@@href[1]{\endgroup#1\@@endlink}%
\providecommand \@sanitize@url [0]{\catcode `\\12\catcode `\$12\catcode `\&12\catcode `\#12\catcode `\^12\catcode `\_12\catcode `\%12\relax}%
\providecommand \@@startlink[1]{}%
\providecommand \@@endlink[0]{}%
\providecommand \url  [0]{\begingroup\@sanitize@url \@url }%
\providecommand \@url [1]{\endgroup\@href {#1}{\urlprefix }}%
\providecommand \urlprefix  [0]{URL }%
\providecommand \Eprint [0]{\href }%
\providecommand \doibase [0]{http://dx.doi.org/}%
\providecommand \selectlanguage [0]{\@gobble}%
\providecommand \bibinfo  [0]{\@secondoftwo}%
\providecommand \bibfield  [0]{\@secondoftwo}%
\providecommand \translation [1]{[#1]}%
\providecommand \BibitemOpen [0]{}%
\providecommand \bibitemStop [0]{}%
\providecommand \bibitemNoStop [0]{.\EOS\space}%
\providecommand \EOS [0]{\spacefactor3000\relax}%
\providecommand \BibitemShut  [1]{\csname bibitem#1\endcsname}%
\let\auto@bib@innerbib\@empty
%</preamble>
\bibitem [{\citenamefont {H\"anggi}\ and\ \citenamefont {Marchesoni}(2009)}]{RMP09}%
  \BibitemOpen
  \bibfield  {author} {\bibinfo {author} {\bibfnamefont {P.}~\bibnamefont {H\"anggi}}\ and\ \bibinfo {author} {\bibfnamefont {F.}~\bibnamefont {Marchesoni}},\ }\href {\doibase 10.1103/RevModPhys.81.387} {\bibfield  {journal} {\bibinfo  {journal} {Rev. Mod. Phys.}\ }\textbf {\bibinfo {volume} {81}},\ \bibinfo {pages} {387} (\bibinfo {year} {2009})}\BibitemShut {NoStop}%
\bibitem [{\citenamefont {Pekola}(2015)}]{pekola_towards_2015}%
  \BibitemOpen
  \bibfield  {author} {\bibinfo {author} {\bibfnamefont {J.~P.}\ \bibnamefont {Pekola}},\ }\href {\doibase 10.1038/nphys3169} {\bibfield  {journal} {\bibinfo  {journal} {Nature Physics}\ }\textbf {\bibinfo {volume} {11}},\ \bibinfo {pages} {118} (\bibinfo {year} {2015})}\BibitemShut {NoStop}%
\bibitem [{\citenamefont {Koski}\ \emph {et~al.}(2014)\citenamefont {Koski}, \citenamefont {Maisi}, \citenamefont {Pekola},\ and\ \citenamefont {Averin}}]{koski_experimental_2014}%
  \BibitemOpen
  \bibfield  {author} {\bibinfo {author} {\bibfnamefont {J.~V.}\ \bibnamefont {Koski}}, \bibinfo {author} {\bibfnamefont {V.~F.}\ \bibnamefont {Maisi}}, \bibinfo {author} {\bibfnamefont {J.~P.}\ \bibnamefont {Pekola}}, \ and\ \bibinfo {author} {\bibfnamefont {D.~V.}\ \bibnamefont {Averin}},\ }\href {\doibase 10.1073/pnas.1406966111} {\bibfield  {journal} {\bibinfo  {journal} {Proceedings of the National Academy of Sciences}\ }\textbf {\bibinfo {volume} {111}},\ \bibinfo {pages} {13786} (\bibinfo {year} {2014})}\BibitemShut {NoStop}%
\bibitem [{\citenamefont {Wagner}\ \emph {et~al.}(2019)\citenamefont {Wagner}, \citenamefont {Talkner}, \citenamefont {Bayer}, \citenamefont {Rugeramigabo}, \citenamefont {Hänggi},\ and\ \citenamefont {Haug}}]{wagner_quantum_2019}%
  \BibitemOpen
  \bibfield  {author} {\bibinfo {author} {\bibfnamefont {T.}~\bibnamefont {Wagner}}, \bibinfo {author} {\bibfnamefont {P.}~\bibnamefont {Talkner}}, \bibinfo {author} {\bibfnamefont {J.~C.}\ \bibnamefont {Bayer}}, \bibinfo {author} {\bibfnamefont {E.~P.}\ \bibnamefont {Rugeramigabo}}, \bibinfo {author} {\bibfnamefont {P.}~\bibnamefont {Hänggi}}, \ and\ \bibinfo {author} {\bibfnamefont {R.~J.}\ \bibnamefont {Haug}},\ }\href {\doibase 10.1038/s41567-018-0412-5} {\bibfield  {journal} {\bibinfo  {journal} {Nature Physics}\ }\textbf {\bibinfo {volume} {15}},\ \bibinfo {pages} {330} (\bibinfo {year} {2019})}\BibitemShut {NoStop}%
\bibitem [{\citenamefont {Masuyama}\ \emph {et~al.}(2018)\citenamefont {Masuyama}, \citenamefont {Funo}, \citenamefont {Murashita}, \citenamefont {Noguchi}, \citenamefont {Kono}, \citenamefont {Tabuchi}, \citenamefont {Yamazaki}, \citenamefont {Ueda},\ and\ \citenamefont {Nakamura}}]{Masuyama2018}%
  \BibitemOpen
  \bibfield  {author} {\bibinfo {author} {\bibfnamefont {Y.}~\bibnamefont {Masuyama}}, \bibinfo {author} {\bibfnamefont {K.}~\bibnamefont {Funo}}, \bibinfo {author} {\bibfnamefont {Y.}~\bibnamefont {Murashita}}, \bibinfo {author} {\bibfnamefont {A.}~\bibnamefont {Noguchi}}, \bibinfo {author} {\bibfnamefont {S.}~\bibnamefont {Kono}}, \bibinfo {author} {\bibfnamefont {Y.}~\bibnamefont {Tabuchi}}, \bibinfo {author} {\bibfnamefont {R.}~\bibnamefont {Yamazaki}}, \bibinfo {author} {\bibfnamefont {M.}~\bibnamefont {Ueda}}, \ and\ \bibinfo {author} {\bibfnamefont {Y.}~\bibnamefont {Nakamura}},\ }\href {\doibase 10.1038/s41467-018-03686-y} {\bibfield  {journal} {\bibinfo  {journal} {Nature Communications}\ }\textbf {\bibinfo {volume} {9}} (\bibinfo {year} {2018}),\ 10.1038/s41467-018-03686-y}\BibitemShut {NoStop}%
\bibitem [{\citenamefont {Naghiloo}\ \emph {et~al.}(2018)\citenamefont {Naghiloo}, \citenamefont {Alonso}, \citenamefont {Romito}, \citenamefont {Lutz},\ and\ \citenamefont {Murch}}]{PhysRevLett.121.030604}%
  \BibitemOpen
  \bibfield  {author} {\bibinfo {author} {\bibfnamefont {M.}~\bibnamefont {Naghiloo}}, \bibinfo {author} {\bibfnamefont {J.~J.}\ \bibnamefont {Alonso}}, \bibinfo {author} {\bibfnamefont {A.}~\bibnamefont {Romito}}, \bibinfo {author} {\bibfnamefont {E.}~\bibnamefont {Lutz}}, \ and\ \bibinfo {author} {\bibfnamefont {K.~W.}\ \bibnamefont {Murch}},\ }\href {\doibase 10.1103/PhysRevLett.121.030604} {\bibfield  {journal} {\bibinfo  {journal} {Phys. Rev. Lett.}\ }\textbf {\bibinfo {volume} {121}},\ \bibinfo {pages} {030604} (\bibinfo {year} {2018})}\BibitemShut {NoStop}%
\bibitem [{\citenamefont {Josephson}(1964)}]{JosephsonRMP1964}%
  \BibitemOpen
  \bibfield  {author} {\bibinfo {author} {\bibfnamefont {B.~D.}\ \bibnamefont {Josephson}},\ }\href {\doibase 10.1103/RevModPhys.36.216} {\bibfield  {journal} {\bibinfo  {journal} {Reviews of Modern Physics}\ }\textbf {\bibinfo {volume} {36}},\ \bibinfo {pages} {216} (\bibinfo {year} {1964})}\BibitemShut {NoStop}%
\bibitem [{\citenamefont {Shapiro}(1963)}]{Shapiro1963}%
  \BibitemOpen
  \bibfield  {author} {\bibinfo {author} {\bibfnamefont {S.}~\bibnamefont {Shapiro}},\ }\href {\doibase 10.1103/PhysRevLett.11.80} {\bibfield  {journal} {\bibinfo  {journal} {Physical Review Letters}\ }\textbf {\bibinfo {volume} {11}},\ \bibinfo {pages} {80} (\bibinfo {year} {1963})}\BibitemShut {NoStop}%
\bibitem [{\citenamefont {Hamilton}(2000)}]{hamiltonJosephsonVoltageStandards2000a}%
  \BibitemOpen
  \bibfield  {author} {\bibinfo {author} {\bibfnamefont {C.~A.}\ \bibnamefont {Hamilton}},\ }\href@noop {} {\bibfield  {journal} {\bibinfo  {journal} {Review of Scientific Instruments}\ }\textbf {\bibinfo {volume} {71}} (\bibinfo {year} {2000})}\BibitemShut {NoStop}%
\bibitem [{\citenamefont {Kautz}(1996)}]{KautzRev1996}%
  \BibitemOpen
  \bibfield  {author} {\bibinfo {author} {\bibfnamefont {R.~L.}\ \bibnamefont {Kautz}},\ }\href {\doibase 10.1088/0034-4885/59/8/001} {\bibfield  {journal} {\bibinfo  {journal} {Reports on Progress in Physics}\ }\textbf {\bibinfo {volume} {59}},\ \bibinfo {pages} {935} (\bibinfo {year} {1996})}\BibitemShut {NoStop}%
\bibitem [{\citenamefont {Kalantre}\ \emph {et~al.}(2020)\citenamefont {Kalantre}, \citenamefont {Yu}, \citenamefont {Wei}, \citenamefont {Watanabe}, \citenamefont {Taniguchi}, \citenamefont {Hernandez-Rivera}, \citenamefont {Amet},\ and\ \citenamefont {Williams}}]{PhysRevResearch.2.023093}%
  \BibitemOpen
  \bibfield  {author} {\bibinfo {author} {\bibfnamefont {S.~S.}\ \bibnamefont {Kalantre}}, \bibinfo {author} {\bibfnamefont {F.}~\bibnamefont {Yu}}, \bibinfo {author} {\bibfnamefont {M.~T.}\ \bibnamefont {Wei}}, \bibinfo {author} {\bibfnamefont {K.}~\bibnamefont {Watanabe}}, \bibinfo {author} {\bibfnamefont {T.}~\bibnamefont {Taniguchi}}, \bibinfo {author} {\bibfnamefont {M.}~\bibnamefont {Hernandez-Rivera}}, \bibinfo {author} {\bibfnamefont {F.}~\bibnamefont {Amet}}, \ and\ \bibinfo {author} {\bibfnamefont {J.~R.}\ \bibnamefont {Williams}},\ }\href {\doibase 10.1103/PhysRevResearch.2.023093} {\bibfield  {journal} {\bibinfo  {journal} {Phys. Rev. Research}\ }\textbf {\bibinfo {volume} {2}},\ \bibinfo {pages} {023093} (\bibinfo {year} {2020})}\BibitemShut {NoStop}%
\bibitem [{\citenamefont {Larson}\ \emph {et~al.}(2020)\citenamefont {Larson}, \citenamefont {Zhao}, \citenamefont {Arnault}, \citenamefont {Wei}, \citenamefont {Seredinski}, \citenamefont {Li}, \citenamefont {Watanabe}, \citenamefont {Taniguchi}, \citenamefont {Amet},\ and\ \citenamefont {Finkelstein}}]{Larson2020}%
  \BibitemOpen
  \bibfield  {author} {\bibinfo {author} {\bibfnamefont {T.~F.~Q.}\ \bibnamefont {Larson}}, \bibinfo {author} {\bibfnamefont {L.}~\bibnamefont {Zhao}}, \bibinfo {author} {\bibfnamefont {E.~G.}\ \bibnamefont {Arnault}}, \bibinfo {author} {\bibfnamefont {M.-T.}\ \bibnamefont {Wei}}, \bibinfo {author} {\bibfnamefont {A.}~\bibnamefont {Seredinski}}, \bibinfo {author} {\bibfnamefont {H.}~\bibnamefont {Li}}, \bibinfo {author} {\bibfnamefont {K.}~\bibnamefont {Watanabe}}, \bibinfo {author} {\bibfnamefont {T.}~\bibnamefont {Taniguchi}}, \bibinfo {author} {\bibfnamefont {F.}~\bibnamefont {Amet}}, \ and\ \bibinfo {author} {\bibfnamefont {G.}~\bibnamefont {Finkelstein}},\ }\href {\doibase 10.1021/acs.nanolett.0c01598} {\bibfield  {journal} {\bibinfo  {journal} {Nano Letters}\ }\textbf {\bibinfo {volume} {20}},\ \bibinfo {pages} {6998} (\bibinfo {year} {2020})}\BibitemShut {NoStop}%
\bibitem [{\citenamefont {Tinkham}(1996)}]{Tinkham}%
  \BibitemOpen
  \bibfield  {author} {\bibinfo {author} {\bibfnamefont {M.}~\bibnamefont {Tinkham}},\ }\href@noop {} {\emph {\bibinfo {title} {Introduction to {{Superconductivity}}}}},\ \bibinfo {edition} {2nd}\ ed.\ (\bibinfo  {publisher} {{Dover}},\ \bibinfo {year} {1996})\BibitemShut {NoStop}%
\bibitem [{\citenamefont {Chan}\ and\ \citenamefont {Stambaugh}(2007)}]{Chan2007}%
  \BibitemOpen
  \bibfield  {author} {\bibinfo {author} {\bibfnamefont {H.~B.}\ \bibnamefont {Chan}}\ and\ \bibinfo {author} {\bibfnamefont {C.}~\bibnamefont {Stambaugh}},\ }\href {\doibase 10.1103/PhysRevLett.99.060601} {\bibfield  {journal} {\bibinfo  {journal} {Phys. Rev. Lett.}\ }\textbf {\bibinfo {volume} {99}},\ \bibinfo {pages} {060601} (\bibinfo {year} {2007})}\BibitemShut {NoStop}%
\bibitem [{\citenamefont {Ricci}\ \emph {et~al.}(2017)\citenamefont {Ricci}, \citenamefont {Rica}, \citenamefont {Spasenovi{\'{c}}}, \citenamefont {Gieseler}, \citenamefont {Rondin}, \citenamefont {Novotny},\ and\ \citenamefont {Quidant}}]{Ricci2017}%
  \BibitemOpen
  \bibfield  {author} {\bibinfo {author} {\bibfnamefont {F.}~\bibnamefont {Ricci}}, \bibinfo {author} {\bibfnamefont {R.~A.}\ \bibnamefont {Rica}}, \bibinfo {author} {\bibfnamefont {M.}~\bibnamefont {Spasenovi{\'{c}}}}, \bibinfo {author} {\bibfnamefont {J.}~\bibnamefont {Gieseler}}, \bibinfo {author} {\bibfnamefont {L.}~\bibnamefont {Rondin}}, \bibinfo {author} {\bibfnamefont {L.}~\bibnamefont {Novotny}}, \ and\ \bibinfo {author} {\bibfnamefont {R.}~\bibnamefont {Quidant}},\ }\href {\doibase 10.1038/ncomms15141} {\bibfield  {journal} {\bibinfo  {journal} {Nature Communications}\ }\textbf {\bibinfo {volume} {8}},\ \bibinfo {pages} {15141} (\bibinfo {year} {2017})}\BibitemShut {NoStop}%
\bibitem [{\citenamefont {Singh}\ \emph {et~al.}(2016)\citenamefont {Singh}, \citenamefont {Peltonen}, \citenamefont {Khaymovich}, \citenamefont {Koski}, \citenamefont {Flindt},\ and\ \citenamefont {Pekola}}]{singh_distribution_2016}%
  \BibitemOpen
  \bibfield  {author} {\bibinfo {author} {\bibfnamefont {S.}~\bibnamefont {Singh}}, \bibinfo {author} {\bibfnamefont {J.~T.}\ \bibnamefont {Peltonen}}, \bibinfo {author} {\bibfnamefont {I.~M.}\ \bibnamefont {Khaymovich}}, \bibinfo {author} {\bibfnamefont {J.~V.}\ \bibnamefont {Koski}}, \bibinfo {author} {\bibfnamefont {C.}~\bibnamefont {Flindt}}, \ and\ \bibinfo {author} {\bibfnamefont {J.~P.}\ \bibnamefont {Pekola}},\ }\href {\doibase 10.1103/PhysRevB.94.241407} {\bibfield  {journal} {\bibinfo  {journal} {Physical Review B}\ }\textbf {\bibinfo {volume} {94}},\ \bibinfo {pages} {241407} (\bibinfo {year} {2016})}\BibitemShut {NoStop}%
\bibitem [{\citenamefont {Spagnolo}\ \emph {et~al.}(2004)\citenamefont {Spagnolo}, \citenamefont {Agudov},\ and\ \citenamefont {Dubkov}}]{Spagnolo_rev}%
  \BibitemOpen
  \bibfield  {author} {\bibinfo {author} {\bibfnamefont {B.}~\bibnamefont {Spagnolo}}, \bibinfo {author} {\bibfnamefont {N.}~\bibnamefont {Agudov}}, \ and\ \bibinfo {author} {\bibfnamefont {A.}~\bibnamefont {Dubkov}},\ }\href@noop {} {\bibfield  {journal} {\bibinfo  {journal} {Acta Physica Polonica B}\ }\textbf {\bibinfo {volume} {35}} (\bibinfo {year} {2004})}\BibitemShut {NoStop}%
\bibitem [{\citenamefont {De~Cecco}\ \emph {et~al.}(2016)\citenamefont {De~Cecco}, \citenamefont {Le~Calvez}, \citenamefont {Sac{\'e}p{\'e}}, \citenamefont {Winkelmann},\ and\ \citenamefont {Courtois}}]{DeCecco2016}%
  \BibitemOpen
  \bibfield  {author} {\bibinfo {author} {\bibfnamefont {A.}~\bibnamefont {De~Cecco}}, \bibinfo {author} {\bibfnamefont {K.}~\bibnamefont {Le~Calvez}}, \bibinfo {author} {\bibfnamefont {B.}~\bibnamefont {Sac{\'e}p{\'e}}}, \bibinfo {author} {\bibfnamefont {C.~B.}\ \bibnamefont {Winkelmann}}, \ and\ \bibinfo {author} {\bibfnamefont {H.}~\bibnamefont {Courtois}},\ }\href {\doibase 10.1103/PhysRevB.93.180505} {\bibfield  {journal} {\bibinfo  {journal} {Physical Review B}\ }\textbf {\bibinfo {volume} {93}} (\bibinfo {year} {2016}),\ 10.1103/PhysRevB.93.180505}\BibitemShut {NoStop}%
\bibitem [{\citenamefont {Clarke}\ \emph {et~al.}(1988)\citenamefont {Clarke}, \citenamefont {Cleland}, \citenamefont {Devoret}, \citenamefont {Esteve},\ and\ \citenamefont {Martinis}}]{Clarke1988}%
  \BibitemOpen
  \bibfield  {author} {\bibinfo {author} {\bibfnamefont {J.}~\bibnamefont {Clarke}}, \bibinfo {author} {\bibfnamefont {A.~N.}\ \bibnamefont {Cleland}}, \bibinfo {author} {\bibfnamefont {M.~H.}\ \bibnamefont {Devoret}}, \bibinfo {author} {\bibfnamefont {D.}~\bibnamefont {Esteve}}, \ and\ \bibinfo {author} {\bibfnamefont {J.~M.}\ \bibnamefont {Martinis}},\ }\href {\doibase 10.1126/science.239.4843.992} {\bibfield  {journal} {\bibinfo  {journal} {Science}\ }\textbf {\bibinfo {volume} {239}},\ \bibinfo {pages} {992} (\bibinfo {year} {1988})}\BibitemShut {NoStop}%
\bibitem [{\citenamefont {Fulton}\ and\ \citenamefont {Dunkleberger}(1974)}]{fultonLifetimeZerovoltageState1974a}%
  \BibitemOpen
  \bibfield  {author} {\bibinfo {author} {\bibfnamefont {T.~A.}\ \bibnamefont {Fulton}}\ and\ \bibinfo {author} {\bibfnamefont {L.~N.}\ \bibnamefont {Dunkleberger}},\ }\href {\doibase 10.1103/PhysRevB.9.4760} {\bibfield  {journal} {\bibinfo  {journal} {Physical Review B}\ }\textbf {\bibinfo {volume} {9}},\ \bibinfo {pages} {4760} (\bibinfo {year} {1974})}\BibitemShut {NoStop}%
\end{thebibliography}%
\newpage
\clearpage
\widetext

\renewcommand{\thepage}{S\arabic{page}}
\renewcommand{\thefigure}{S\arabic{figure}}
\renewcommand{\thesection}{S\arabic{section}}
\setcounter{figure}{0}
\setcounter{page}{1}
\setcounter{equation}{0}

\begin{center}
\textbf{\huge Supplementary Materials}
\end{center}

\section{Measurement Details}

In order to minimize distortions, all presented switching data in Figures 1-3 correspond to switching rates slower than 1 kHz, which is 5 times slower than our sampling rate and 10 times slower than the cutoff of the cryogenic low pass filters.  All data is also inspected to make sure less than 5$\%$ lies between the voltages $\pm \frac{1}{2} \frac{hf}{2e}$ corresponding to one half of the quantized values.

Reported values of $\tau_{0}$ are measured from  the $\sqrt{\tau_{+}\tau_{-}}$ (Figure~2a). This helps to correct for small bias imbalances, to which the system is particularly sensitive at elevated temperatures. 

\begin{figure*}[b]
    \includegraphics[scale = 0.4]{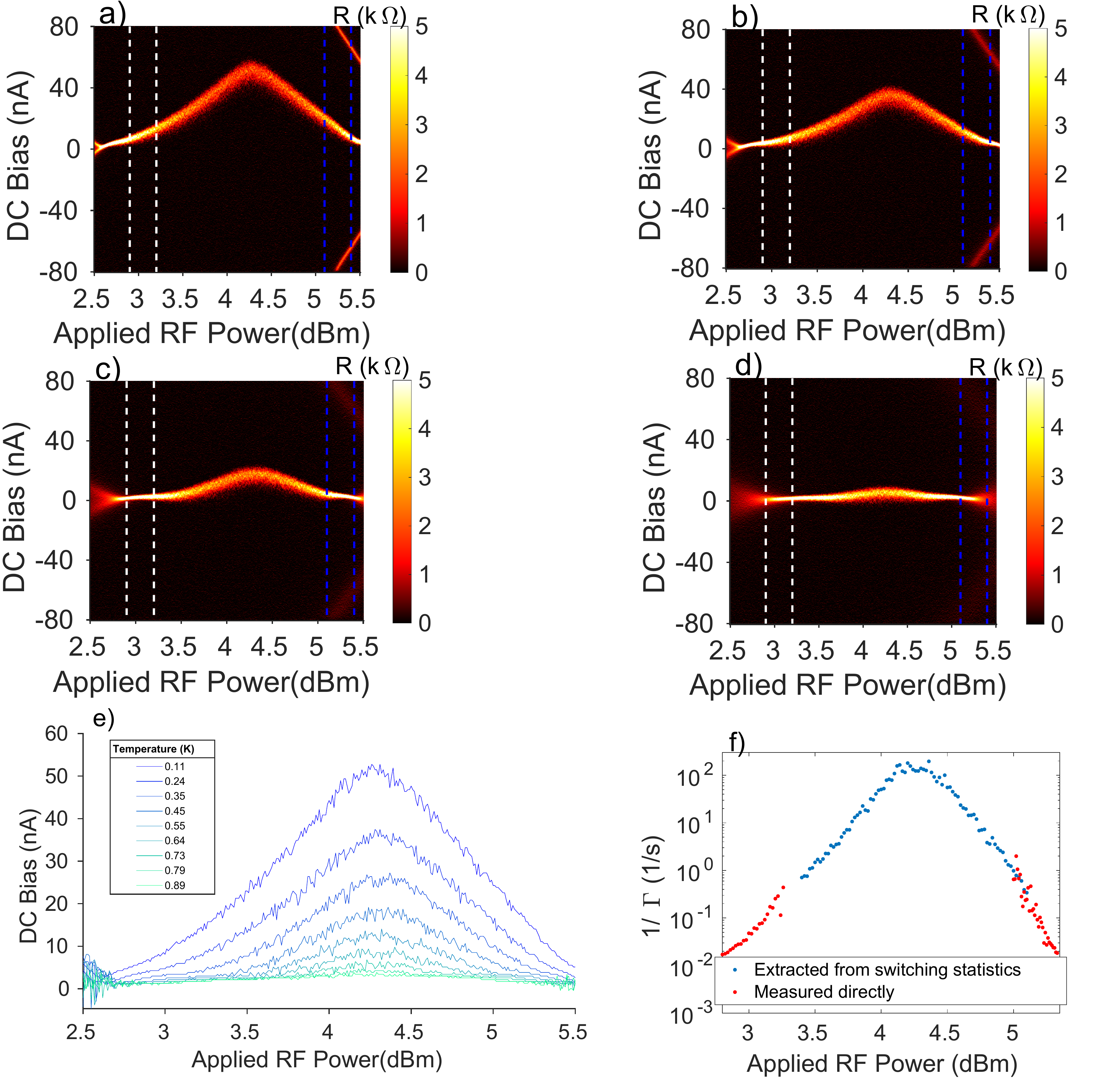}
\caption{a-d) Evolution of the hysteretic  boundary between $n=1$ and $n=-1$ steps measured at temperatures of 100, 250, 450 and 730 mK, respectively. These panels show that the hysteresis decreases with temperature, as expected, and that the RF power width of the zero crossing regime gradually decreases as temperature increases.  This helps to rule out any trivial mechanism for the observed non-monotonicity, as it should lead to a shorter lifetime at high temperature rather than a longer one.  e) The switching current is extracted from (a-d) and other intermediate temperatures.  Lastly, in panel (f) represents the lifetime $\tau_0$ at zero DC bias and the base temperature across the entire range of $P_{RF}$.  Red dots are measured directly, while blue dots are extracted from the distribution of the switching currents.} 
\end{figure*}

\section{Evolution of the $-1 \rightarrow +1$ Boundary with Temperature}
Figure~1b demonstrates a pronounced hysteresis depending on the bias sweep direction, which results in the arching of the boundary between the $n=\pm1$ states. Several copies of this map taken at different temperatures are displayed in Figure~S1a-d. As one might expect, the switching hysteresis is suppressed at higher temperatures, eventually disappearing above 1 K. A summary containing the extracted switching currents at different temperatures is shown in Figure~S1e. 

It is also possible to plot the transition rate across the entire zero crossing regime with minor caveats (Figure~S1f). In the middle of the zero crossing regime (RF power between 3.5 and 5 dB), the very long lifetime makes it difficult to measure the switching directly.  Instead, we can perform measurements of the switching current distribution while ramping the DC bias current.  This distribution can be used to infer $\Gamma(I)$ for $I \lesssim I_{C}$\cite{fultonLifetimeZerovoltageState1974a}.  Given that $\Gamma(I)$ follows an exponential fit in this range, it is possible to extrapolate to $\Gamma(0)$. This procedure warrents caution, owing to the fact that we are using data of $\Gamma(I)$ for $I \approx I_{C}$ to extrapolate down to $I = 0$. However, we find that it generates a good match with the directly measured switching time -- see the overlap between the red and blue symbols at RF power around 5dB in Figure~S1f.

It is worth noting that some care must be taken when relating $\tau$, extracted from direct measurements of the bistable regime, and $\Gamma$, which is extracted by measuring the switching current distribution.  Attempting to stitch together data from the two different methods using $\gamma = \frac{1}{\tau}$ produces a significant disagreement with the methodologies, presumably due to the fact that, generically, $\tau \neq 1/( \int_{0}^{\infty}\frac{1}{\tau'} P(\tau')d\tau')$.  To allow for the different datasets to stitch properly, we directly average $\frac{1}{\tau}$.

\section{Elimination of the Trivial Explanations for the Non-Monotonic $\tau_0(T)$}

\begin{figure}[b]
    \includegraphics[scale = 0.4]{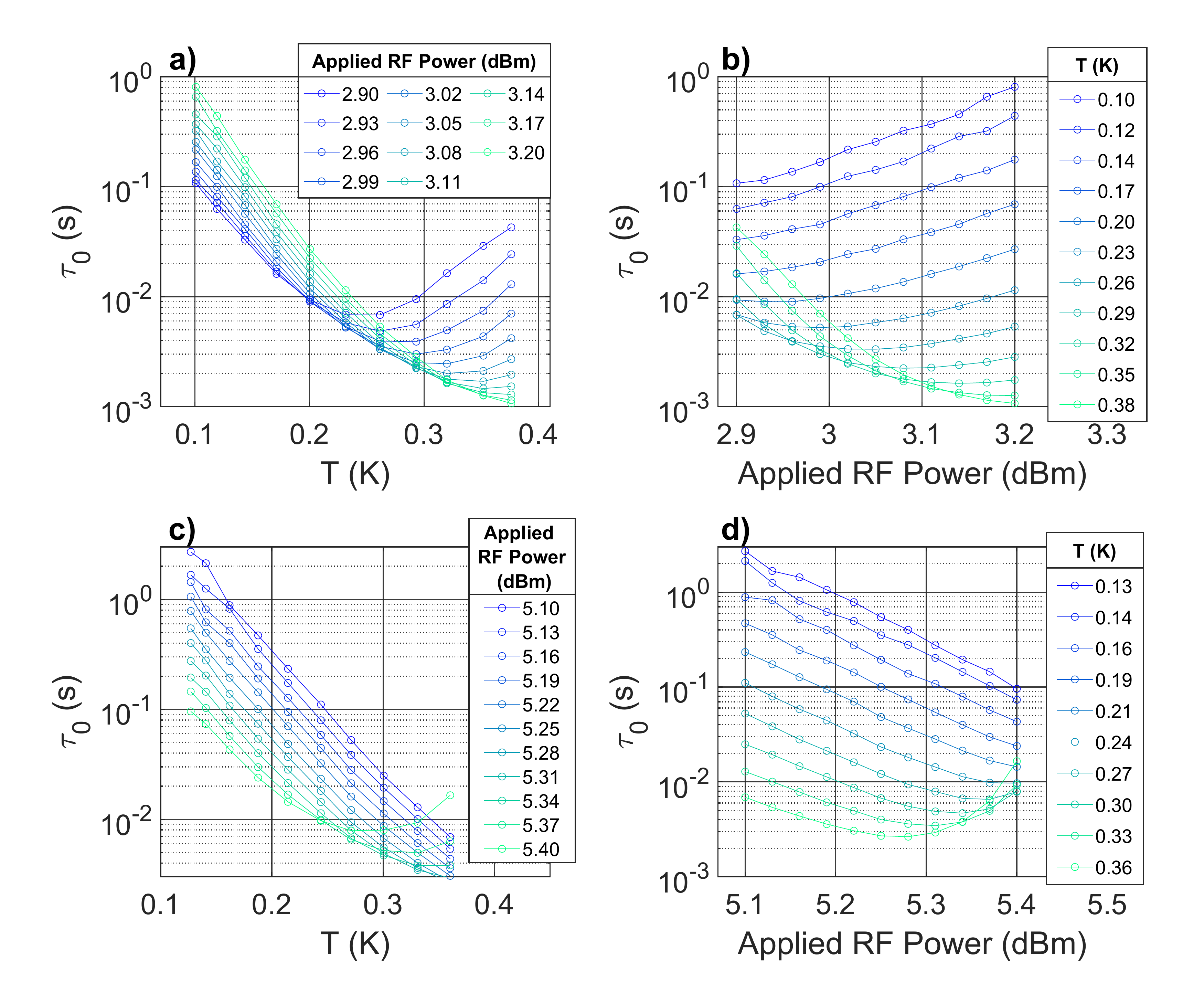}
\caption{ Switching times $\tau_0(T,P)$, plotted vs temperature at fixed powers (left) and vs power at fixed temperatures (right). The top row is measured close to the first bifurcation point ($P_{RF}$ between 2.9 and 3.1 dBm). Panel (a) is a reproduction of Figure~2b. In the bottom row, $\tau_0(T,P)$ is measured close to the second bifurcation point at which the $n=0$ step reappears ($P_{RF}$ between 5.1 and 5.4 dBm). While the same characteristic behavior is observed, the non-monotonic $\tau_0(T)$ is first observed in panel (a) at the lower range of powers and in panel (c) at the higher range of powers. As discussed in the text, this observation allows us to eliminate some trivial scenarios.}
\end{figure}

We have considered and discarded several possible alternative explanations for the nonmonotonic dependence of $\tau_0(T)$, primarily related to unintended shifts in control parameters. One such argument is that increasing the temperature may decrease $I_{C}$, which would move the zero-crossing regime in Figure~1a toward lower RF power levels. While Figure~S1 presents evidence that the boundary actually shifts in the opposite direction with temperature, such that this should cause $\tau_{0}$ to further decrease, we also present additional experimental evidence addressing this concern. 

To this end, we perform measurements close to the second bifurcation point, in the range of RF powers marked by blue lines in Figure~S1. (The bifurcation point itself occurs around 5.6 dBm, just to the right of the frames of Figure~S1.) The measured switching times near the second bifurcation point are shown on the bottom row of Figure~S2. To facilitate the direct comparison, the data measured near the first bifurcation point is plotted in the top row of Figure~S2. (This set of data was shown in Figure~2b of the main text and corresponds to the range of RF power marked by white lines in Figure~S1.) For both sets, the same data is plotted vs. temperature (left column) and vs. RF power (right column). 

For the second bifurcation point, we find that non-monotonicity in temperature first sets in at the \emph{highest power} (Figure~S2c). This is opposite to the first bifurcation point, where it occurs first at the \emph{lowest power} (Figure~S2a). This observation indicates that the key parameter controlling the non-monotonicity is not the total power, but the proximity to the bifurcation point. However, the relation of the non-monotonicity and the proximity to the bifurcation is subtle: the bifurcation points move slightly closer together as the temperature increases, see Figure~S1a-d. Naively, this could only shorten the switching time, contrary to our observations.

The observations of Figure~S2 also rule out many trivial explanations related to additional parameters changing as a function of temperature. For example, both bifurcation points would move to the lower RF powers as the critical current is decreased at higher temperatures; as a result both would show non-monotonicity at the lowest powers, contrary to our observations.

\section{Behaviour of $\tau_0$ at the Higher Temperatures}

The non-monotonic curves in Figure~S2 end with the switching time increasing at the high temperature end. However, this enhanced stability might be expected to eventually reverse at even higher temperatures. In practice, as the step voltage values start to decrease at higher temperatures, the time trace include more values which are not quantized, which makes it more difficult to define the two states $n = \pm 1$. However, there were a few parameter regimes for which it was possible to observe a decrease in $\tau_{0}$ at the high temperature, while the data still remained quite bimodal. One example of such behavior is shown in Figure~S3. 

\begin{figure}[h]
    \includegraphics[scale = 0.45]{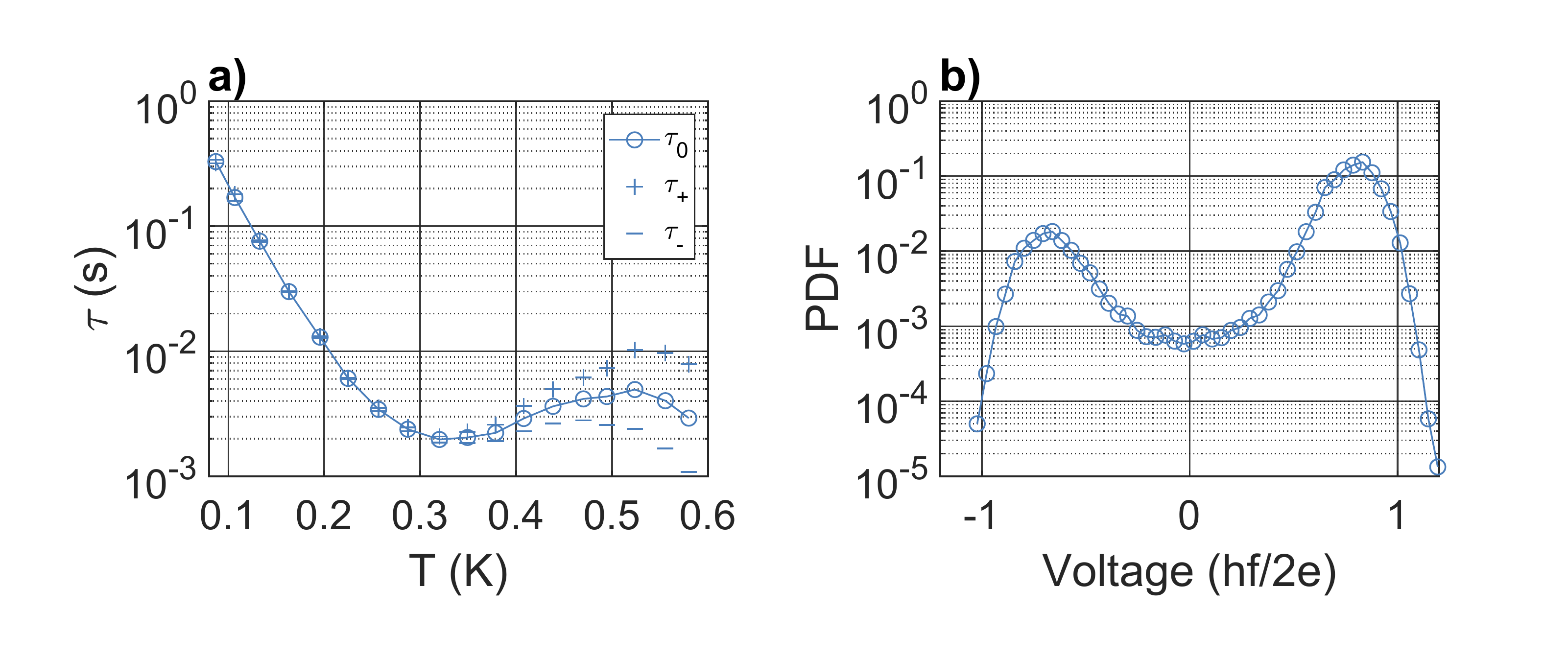}
\caption{a) $\tau_0$, $\tau_{+}$, and $\tau_{-}$ at $P_{RF}=3.14$ dBm (the third green curve in Figure~S2a) shown in the extended temperature range. The eventual drop of the switching times at the highest temperatures is visible in all three quantities. Note that a small unintentional current offset results in the difference between $\tau_{+}$ and $\tau_{-}$ which becomes more apparent at higher temperatures. However, the downturn is apparent in both $\tau_{+}$ and $\tau_{-}$, and the shortest $\tau_{-}$ is still 5 times larger than the sampling time scale, which allows us to extract $\tau_0 = (\tau_{+}\tau_{-})^{1/2}$ (see Figure~2a). b) The histogram corresponding to the highest temperature data point in (a). It shows that the distribution of voltages is still bimodal at this temperature (note the logarithmic scale), although both steps are reduced compared to the quantized values, presumably due to the rapid excursions to the $n=0$ state.}
\end{figure}

\begin{figure}
    \includegraphics[scale=0.4]{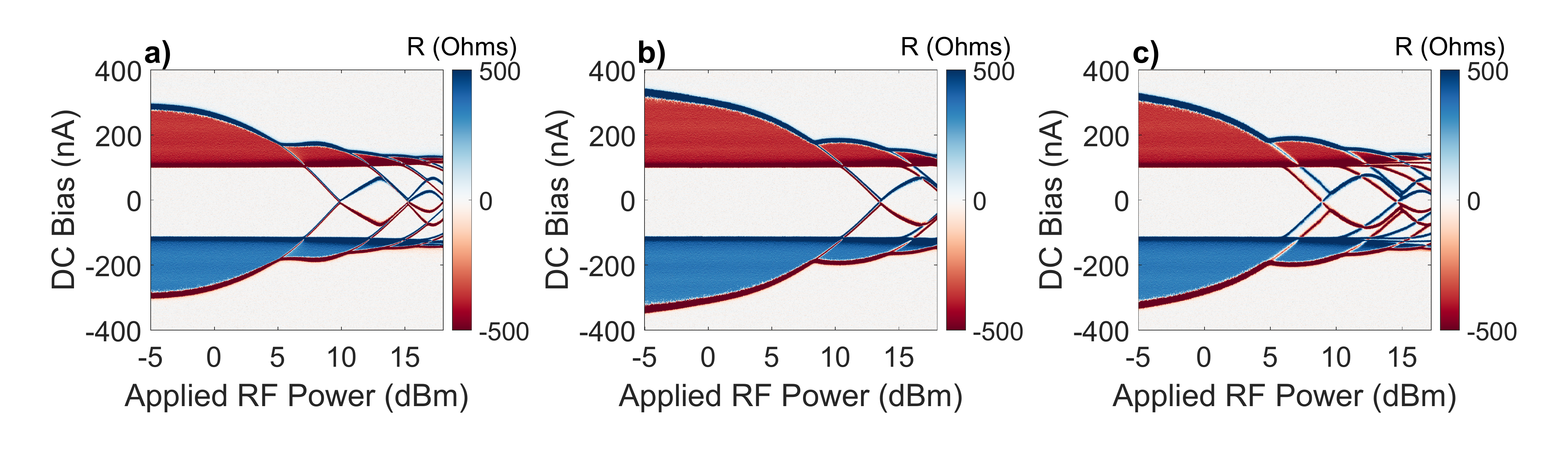}
\caption{The symmetrized Shapiro maps for (a) 5.4 GHz, (b) 5.9 GHz, and (c) 6.42 GHz. The maps represent the difference between the $dV/dI$ data measured in the opposite sweep directions. Such maps highlight both the hysteresis between the sweep directions and the overall asymmetry. The features that appear doubled in these maps are hysteretic. At 5.4 GHz, the $\pm 1 \rightarrow 0$ and $0 \rightarrow \pm 1$ boundaries almost coincide, while at 6.4 GHz significant hysteresis develops, pointing to a greater stability of the $n=0$ plateau. 
}
\end{figure}

\section{Tristable behavior at high frequency }

In this section, we directly reveal the role of the $n=0$ state in the system dynamics. We start by exploring the coexistence region between the $n=0$ and $\pm 1$ states. 

The Shapiro maps of Figures 1 and S1 are highly hysteretic. We can highlight the hysteresis by subtracting the maps measured in the two sweep directions. After correcting for any delay caused by the cryogenic filters, nonhystereic features will overlap and nearly cancel, while the hysteretic features become doubled. 

In Figure~S4, we show such symmetrized maps measured in a range of frequencies from 5.4 to 6.42 GHz. 
We focus on the boundaries between $n=0$ and $n=\pm 1$ states at the closing of the $n=0$ step. At 5.4 GHz the boundaries nearly coincide, while for 6.42 GHz they are well separated. As a result, at 6.42 GHz, the boundaries form three separate intersection points. The original bifurcation point appears at the intersection of the $-1 \rightarrow 0$ and $+1 \rightarrow 0$ boundaries at zero bias. Two more intersections points appear symmetrically at the positive and negative bias, for example one of them marks the intersection of the $-1 \rightarrow 0$, $0 \rightarrow +1$, and $-1 \rightarrow +1$ boundaries.

Since the sweep rate and noise are constant throughout the measurement, we take the appearance of the hysteresis as an indication that the quasipotential profile has grown substantially deeper at 6.42 GHz. We argue that at lower RF frequencies, such as 5 and 5.4 GHz, as the stability of $n=+1,-1$ is increased with applied RF power, the $n=0$ becomes unstable on the measurement scales. On the contrary, Figure~S3d indicates that at 6.42 GHz all three states have comparable stability. We next measure the voltage traces at this drive frequency in the power range between these three intersection points, Figure~S5a. We indeed find that all three voltages states can be simultaneously stable, as is shown in Figure~S5b. 

We note that this regime is quite distinct from the non-Markovian regime explored in Figure~4 of the main paper and modeled in Section S6. Here, all transitions appear to be Markovian -- careful inspection of the full version (600 sec.) of the trace in Figure~S4a indicates that the system is equally likely to switch from $n=0$ to either $n=1$ or $n=-1$ regardless of the history. This is to be expected, because our assumption is that the memory provided by the RC circuitry dissipates on a scale shorter than our measurement time scale; therefore if we can directly observe a robust $n=0$ state, the system cannot demonstrate non-Markovian switching.

Despite this difference between the 6.42 GHz and lower frequencies, there should exist a continuous transition between them. The direct observation of switching via the $n=0$ state lends support to the idea that at lower RF frequencies the system should also evolve via the $n=0$ state, albeit on the time scales which are too short relative to the measurement bandwidth.

\begin{figure}
    \includegraphics[scale = 0.45]{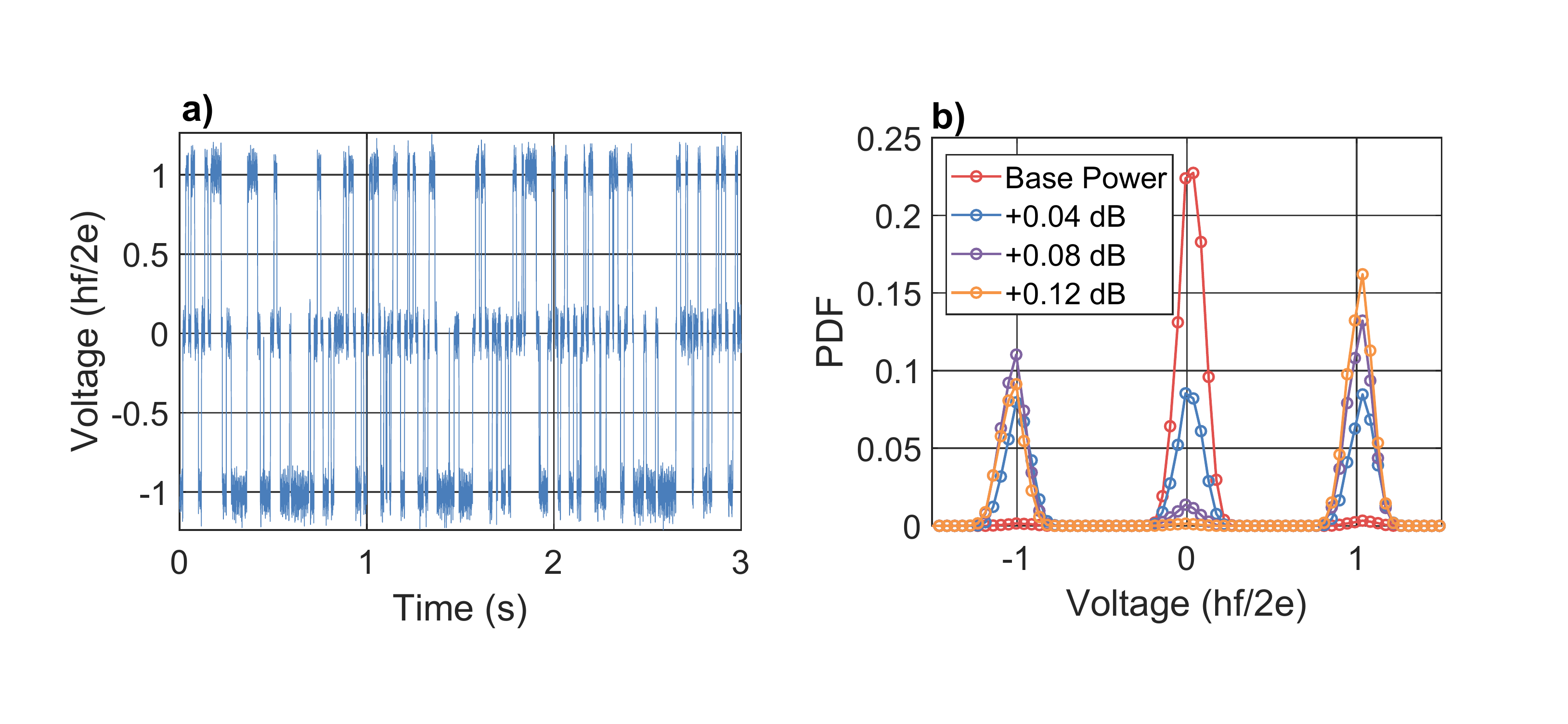}
\caption{Resolvable tristable data measured with 6.42 GHz RF drive. a) A sample time trace, showing the presence of all 3 stable voltages.  There are essentially no direct transitions between $n=-1$ and $n=+1$, and instead all transitions pass through $n=0$. The applied power corresponds to +0.04 dB in the next panel b) The distribution of observed voltages, plotted for a range of different applied RF powers. At lowest power, $n=0$ is observed almost exclusively (blue curve); at the intermediate power all three states have the same probability (red); and at the highest power, $n=0$ is significantly suppressed (yellow). Note that at 5 GHz it was not possible to achieve relative balance between all three steps (as in the red curve). Presumably, in that case the switching rate greatly exceeded the measurement bandwidth.}
\end{figure}

\section{Full Numerical Model}
We have previously explored the Shapiro maps directly simulating the dynamics of the circuit model in Ref.~\cite{Larson2020}. Here we extend this simulation with an additional RC stage attached, Figure~S4a. This leads to a straightforward extension of the standard Shapiro step equations:

\begin{equation}
    \begin{aligned}
        I&=I_{bias}+I_{RF} \sin{\omega t}+I_N(t)\\
        &=C_{1} \frac{dV_{1}}{dt}+C_0 \frac{dV_{0}}{dt}+ I_C \sin{\phi} +\frac{\hbar}{2 e R_j}\frac{d\phi}{dt}+\frac{\hbar C_j}{2 e}\frac{d^2 \phi}{dt^2}\\
        V_{0}&=\frac{\hbar}{2e}\frac{d\phi}{dt}+R_L\left(I_C \sin{\phi} +\frac{\hbar}{2 e R_j}\frac{d\phi}{dt}+\frac{\hbar C_j}{2 e}\frac{d^2 \phi}{dt^2}\right)\\
        V_{1}&=\frac{\hbar}{2e}\frac{d\phi}{dt}+(R_L+R_1)\left(I_C \sin{\phi} +\frac{\hbar}{2 e R_j}\frac{d\phi}{dt}+\frac{\hbar C_j}{2 e}\frac{d^2 \phi}{dt^2}\right)\\
        &+R_1C_0 \frac{dV_0}{dt}
    \end{aligned}
    \label{RCSJ:eq1}
\end{equation}

where $V_{0}$ is the voltage on capacitor $C_{0}$ and $V_{1}$ is the voltage on capacitor $C_{1}$.

The large separation of time scales makes it overly time consuming to simulate the non-monotonic behavior of $\tau_0(T)$. However, we find it fairly straightforward to recreate the correlated switching behavior of Figure~4 of the main text. A portion of one such time trace is shown in Figure~S6. Here, we make use of physically reasonable parameters in Table 1.  
\begin{table}[]
\begin{tabular}{ll}
\hline
\multicolumn{1}{|l|}{Parameter} & \multicolumn{1}{l|}{Value}    \\ \hline
\multicolumn{1}{|l|}{$I_{C}$}   & \multicolumn{1}{l|}{600 nA}   \\ \hline
\multicolumn{1}{|l|}{$R_{J}$}   & \multicolumn{1}{l|}{300 Ohm}  \\ \hline
\multicolumn{1}{|l|}{$R_{0}$}   & \multicolumn{1}{l|}{50 Ohm}   \\ \hline
\multicolumn{1}{|l|}{$C_{0}$}   & \multicolumn{1}{l|}{1.5 pF}   \\ \hline
\multicolumn{1}{|l|}{$R_{1}$}   & \multicolumn{1}{l|}{1500 Ohm} \\ \hline
\multicolumn{1}{|l|}{$C_{1}$}   & \multicolumn{1}{l|}{100 pF}   \\ \hline
\multicolumn{1}{|l|}{$\sigma$}  & \multicolumn{1}{l|}{0.6}      \\ \hline
\multicolumn{1}{|l|}{$f_{AC}$}  & \multicolumn{1}{l|}{5 GHz}      \\ \hline
\multicolumn{1}{|l|}{$I_{AC}$}  & \multicolumn{1}{l|}{1.85 $\mu$A}      \\ \hline 
\end{tabular}
\caption{Values used for the full numerical simulation in Figure~S6. Here, $\sigma$ represents the amplitude [Standard Deviation?] of the $I_{N}(t)$ term in units of $I_{C}$. While this amplitude may seem large, in our model it is applied on the opposite side of the filter as the junction, necessitating its large amplitude. Further, the simulations have to be performed with relatively large noise to allow for relatively fast switching enabling timely computation. 
}
\end{table}

\begin{figure}
    \includegraphics[scale = 0.45]{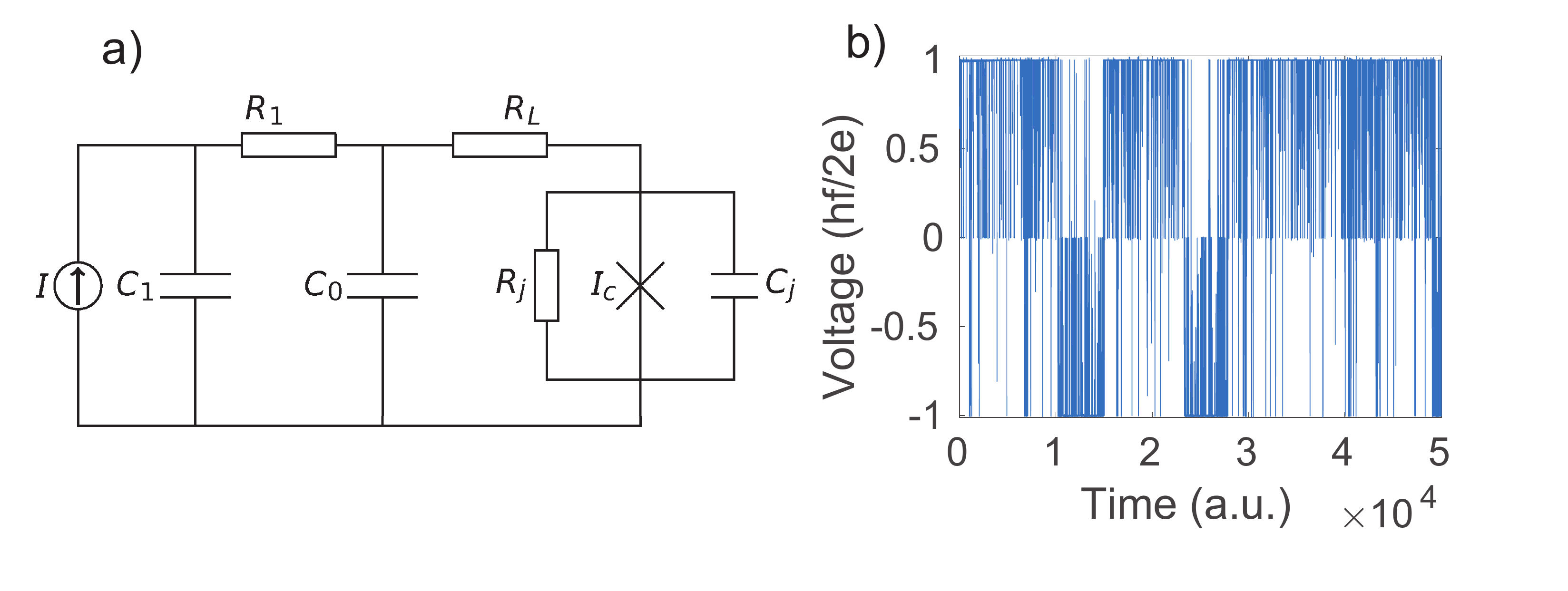}
\caption{a) Electrical circuit used to simulate the effect of the external RC filter. It is analogous to the circuit we previously used in Ref.~\cite{Larson2020} to simulate Shapiro steps, with an added RC stage. b) A time trace obtained by running the model over many Thousands of RF cycles, and averaging the signal over the last 10 cycles to produce one voltage point (as measured at the junction itself). Notice that for large periods of time, the system rapidly bounces between $n=0$ and one of the Shapiro steps (e.g. $n=+1$), only occasionally making short excursions to the opposite step (in this case, $n=-1$). This clearly indicates that memory is at play: from $n=0$, the system is more likely to go back to the state where it has spent time just beforehand.}
\end{figure}

\section{Toy Model}
In order to determine the minimal essential elements for the observations of non-monotonic $\tau_0(T)$, we consider the following toy model. Inspired by the tristable data (section S4) and the full numerical simulations (section S5), wherein any switching between $n=+1$ and $n=-1$ goes through $n=0$, we represent the system by a particle in a three-well potential.  Without loss of generality, we can assume that the particle starts in the left well (corresponding to $n=-1$), and solve for the average time to reach the right well (corresponding to $n=1$), $\tau_{-1\rightarrow 1}$, which will be taken to represent $\tau_0$. In this section, we will use the notation $\tau_{i\rightarrow j}$ to define the average time to transition from well $i$ to well $j$. 

In a symmetric tristable potential, we have $\tau_{0 \rightarrow -1} = \tau_{0 \rightarrow 1}$ and the particle has a 50$\%$ change to go to the right well after reaching the middle well.  Therefore, the average time to reach the right well will be $\tau_{-1 \rightarrow 1}=2\tau_{-1 \rightarrow 0}$. (Here, we neglect the escape times from $n=0$, the metastable middle well, assuming $\tau_{0 \rightarrow \pm1} \ll \tau_{-1 \rightarrow 0}$.)

Even if the potential is not symmetric, and the system is more likely to escape from the middle well to the left well, we can simply write the expected time to reach the right well as a geometric series made of multiple attempts. The series sums to $\tau_{-1 \rightarrow 1}=\tau_{-1 \rightarrow 0}/P_{0\rightarrow1}$ where $P_{0\rightarrow1}$ is the probability to go from the 0 well to +1. Introducing $\Gamma_{+/-} \equiv 1/\tau_{0 \rightarrow \pm 1}$, the escape rates from $n=0$ to $n=\pm 1$, we find that  $P_{0\rightarrow1} =\frac{\Gamma_{+}}{\Gamma_{+}+\Gamma_{-}}$, which results in $\tau_{-1 \rightarrow 1} = \tau_{-1 \rightarrow 0} (\Gamma_{+}+ \Gamma_{-})/\Gamma_{+}$. As expected, this result does not naively produce non-monotonicity if $\tau_{-1 \rightarrow 0}$ and $\Gamma_{+/-}$ all follow a simple activational dependence on temperature.

For a slightly more complicated model, we consider the effect of an external RC on the escape time.  As mentioned in the main text, such external RCs naturally arise in the cryogenic filters and cables. In full, these filters raise our second order differential equation to 3rd order (or higher, depending on the construction of the filter).  However, the $RC$ time of the filter should be much longer the the time scale of junction dynamics, so we treat it adiabatically as a memory effect on our tristable well picture.

Physically, the RC circuit should be charged by the DC voltage developed in the junction.  As the junction switches from the $n=-1$ voltage to the $n=0$ voltage, the RC will start to discharge, current-biasing the junction. This will increase the probability to return to $n=-1$ and decrease the probability to reach $n=+1$.  We can describe these corresponding switching from the middle well with time-dependent rates $\Gamma_{+}(t)$ and $\Gamma_{-}(t)$. To make the problem most tractable, we will approximate them as time-independent constants $\Gamma_{+} < \Gamma_{-1}$ for $t< t_{0}$, where $t_0$ is taken to represent the $RC$ time of the circuit. After that time, we will assume that the capacitors have completely discharged and the particle will have a 50$\%$ chance of going in either direction. 

\begin{figure}
    \includegraphics[scale = 0.4]{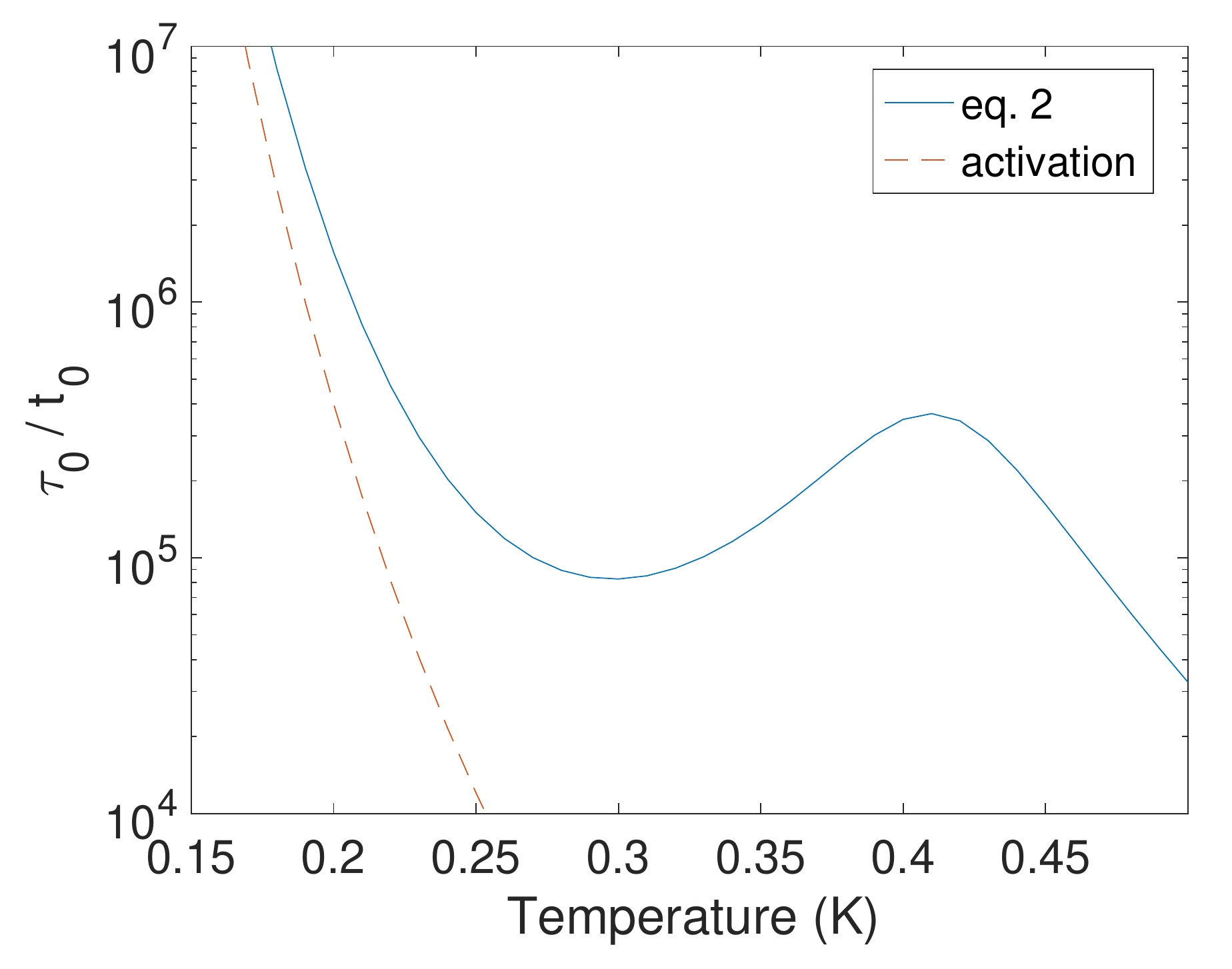}
\caption{Transition time $\tau_0=\tau_{-1 \rightarrow 1}$ between $n=-1$ and $+1$ states obtained from the basic toy model of eq.~(\ref{Toy}).  All the individual transition rates, including $\tau_{-1 \rightarrow 0}$ (red dash), are assumed to follow the standard activational behavior. However, the total $n=-1$ to $n=1$ transition time (blue line) shows a pronounced non-monotonicity. At lower temperatures, it closely follows $\tau_{-1 \rightarrow 0}$. However, it significantly deviates at higher temperatures, owing to the increased rate of return from $n=0$ to $n=-1$. }
\end{figure}

Under these assumptions, it is straightforward to solve for the mean switching time from $n=-1$ to $n=1$:

\begin{equation}
        \tau_{-1 \rightarrow 1} = \frac{\tau_{-1 \rightarrow 0}(\Gamma_{+} + \Gamma_{-})}
        {\Gamma_{+}-\frac{1}{2}(\Gamma_{+}-\Gamma_{-})\exp[-t_0 (\Gamma_{+}+\Gamma_{-})]}
    \label{Toy}
\end{equation}

The two limiting cases discussed earlier can be clearly reproduced. If the mean escape rate from $n=0$ is low, $\Gamma_{+}+\Gamma_{-}\ll 1/t_{0}$ the RC memory has little effect on the system, as the capacitor has fully discharged long before the system switches from $n=0$, and we are back to the symmetric tristable potential picture that we have previously considered. In the opposite limit, $\Gamma_{+}+\Gamma_{-}\gg 1/t_{0}$ the system almost always switches from $n=0$ before the capacitor has discharged, effectively experiencing an asymmetric tristable behavior considered earlier. While the first limit results in $\tau_{-1 \rightarrow 1} = 2\tau_{-1 \rightarrow 0}$, in the second case $\tau_{-1 \rightarrow 1} = \tau_{-1 \rightarrow 0}(\Gamma_{+}+\Gamma_{-})/\Gamma_{+}$, just like above. If $\Gamma_{-} \gg \Gamma_{+}$ the second limit will greatly exceed the first, and a region of non-monotonic $\tau_{-1 \rightarrow 1}$ may appear stitching the two limits together.

Figure~S6 illustrates this nonmonotonic behavior for the set of parameters intended to visually reproduce the experimental behavior. The dashed line represents $\tau_{-1 \rightarrow 0}$, while the solid line represents $\tau_{-1 \rightarrow +1}$. We assume $\tau_{-1 \rightarrow 0} = 10^{-2} t_0 e^{E_A/T}$, $1/\Gamma_{-}=10^{-2} t_0 e^{E_-/T}$ and $1/\Gamma_{+}=10^{-2} t_0 e^{E_+/T}$. Their  temperature-independent energy barriers are taken to be $E_A=3.5$, $E_-=1$, and $E_+=5$. (All energies can be assumed to measured in Kelvin.) 

The intuition for why non-monotonicity arises is as follows. At low temperature, we start with $\Gamma_{+/-}t_0 \ll 1$ and the memory effects are negligible. As the temperature increases and the system escapes from $n=0$ faster and faster, it is more likely to escape before the external RC has discharged, predominantly returning to $n=-1$. The system has to jump back and forth between $n=-1$ and $n=0$ many times before finally transitioning to $n=1$. Thus, even though every individual rate ($1/\tau_{-1 \rightarrow 0}$, $\Gamma_{+/-}$) follows the Arrhenius law, the resulting time to reach $n=+1$ is non-monotonic.

An additional feature of this explanation is that it naturally accounts for the observed reduction in the voltage steps at higher temperatures (see Figure~3 of the main text), as the system averages between $n=-1$ and $n=0$, or between $n=1$ and $n=0$ states.

While the step function discharging model is not realistic, we believe it should still capture all of the desired qualitative features while greatly simplifying calculations.  
We have also assumed that $\tau_{-1\rightarrow 0} \gg t_{0}$ for all temperatures, so that the RC circuit has enough time to reset to $n=-1$ after switching from $n=0$. In other words, we can neglect any memory of the $n=0$ state on $\tau_{-1\rightarrow 0}$. Finally, we neglected the memory effect on switching back from $n=1$ once that state is reached. We know that such memory is present (Figure~4 of the main text), resulting in preferential switching back to $n=0$ and $n=-1$. 
All these effects are important, but we do not think they are necessary for a toy model that describes the nonmonotonic $\tau_0(T)$.

There are several interesting ways to extend the toy model. We can introduce the effects of small finite bias by taking $\Gamma_{0 \rightarrow -1} \neq \Gamma_{0 \rightarrow 1}$ even for $t>t_{0}$ when the RC discharges. We have verified that one can qualitatively capture the bias dependence shown in Figure~2d of the main text. Another direction is to explore the necessity of the $n=0$ well, by simply removing it and introducing the dependence of $\tau_{-1 \rightarrow 1}$ and $\tau_{1 \rightarrow -1}$ on the state of the RC filter. 
This modification of the model may be forced to produce the non-monotonic temperature dependence, indicating that the $n=0$ state may not be strictly essential to generate non-Markovian switching. However, both the experimental data and the full numerical simulations of section S5 demonstrate that the $n=0$ state is essential in our case.

\end{document}